\documentclass[%
 aip,
 amsmath,amssymb,
 reprint,%
]{revtex4-1}
\usepackage{amssymb}
\usepackage{epsfig}
\usepackage{graphicx}
\usepackage{amsmath}
\usepackage{times}
\usepackage{xcolor}
\usepackage{subfig}
\usepackage{caption}

\begin{document}

\title{Comparison of Methods for the Assessment of Nonlinearity in Short-Term Heart Rate Variability under different Physiopathological States}

\author{Luca Faes}
\affiliation{Department of Engineering, University of Palermo, Italy}

\author{Manuel G\'omez-Extremera}
\affiliation{Dpto. de F\'isica Aplicada II, ETSI de Telecomunicaci\'on, University of M\'alaga, 29071 M\'alaga, Spain}

\author{Riccardo Pernice}
\affiliation{Department of Engineering, University of Palermo, Italy}

\author{Pedro Carpena}
\affiliation{Dpto. de F\'isica Aplicada II, ETSI de Telecomunicaci\'on, University of M\'alaga, 29071 M\'alaga, Spain}
\affiliation{Instituto Carlos I de F\'isica Te\'orica y Computacional, University of M\'alaga}

\author{Giandomenico Nollo}
\affiliation{Department of Industrial Engineering, University of Trento, Italy}

\author{Alberto Porta}
\affiliation{Department of Biomedical Sciences for Health, University of Milan, Italy}
\affiliation{Department of Cardiothoracic, Vascular Anesthesia and Intensive Care, IRCCS Policlinico San Donato, San Donato Milanese, Milan, Italy}

\author{Pedro Bernaola-Galv\'an}
\affiliation{Dpto. de F\'isica Aplicada II, ETSI de Telecomunicaci\'on, University of M\'alaga, 29071 M\'alaga, Spain}
\affiliation{Instituto Carlos I de F\'isica Te\'orica y Computacional, University of M\'alaga}

\begin{abstract}
Despite the widespread diffusion of nonlinear methods for heart rate variability (HRV) analysis, the presence and the extent to which nonlinear dynamics contribute to short-term HRV is still controversial. This work aims at testing the hypothesis that different types of nonlinearity can be observed in HRV depending on the method adopted and on the physiopathological state. Two entropy-based measures of time series complexity (normalized complexity index, NCI) and regularity (information storage, IS), and a measure quantifying deviations from linear correlations in a time series (Gaussian linear contrast, GLC), are applied to short HRV recordings obtained in young (Y) and old (O) healthy subjects and in myocardial infarction (MI) patients monitored in the resting supine position and in the upright position reached through head-up tilt. The method of surrogate data is employed to detect the presence of and quantify the contribution of nonlinear dynamics to HRV. We find that the three measures differ both in their variations across groups and conditions and in the number and strength of nonlinear HRV dynamics detected: at rest, IS reveals a significantly lower number of nonlinear dynamics in Y, whereas during tilt GLC reveals significantly stronger nonlinear HRV dynamics in MI; in the transition from rest to tilt, all measures detect a significant weakening of nonlinear HRV dynamics in Y, while only GLC detects a significant strengthening of such dynamics in MI. These results suggest that distinct dynamic structures, detected with different sensitivity by nonlinear measures, lie beneath short-term HRV in different physiological states and pathological conditions.

\end{abstract}




\maketitle

\begin{quotation}
Historically, the study of heart rate variability (HRV) has both received clinical attention, e.g. as a tool for risk stratification after myocardial infarction, and has attracted the interest of physicists who saw it as a particularly lucid example of chaos in physiology. Later on, after it was realized that a thorough evaluation of the chaotic nature of cardiac dynamics is precluded by difficulties inherent in the noisy nature of biological signals and in the restricted length of the data typically available, the field of HRV analysis underwent a shift in paradigm from chaos to complexity and nonlinear dynamics assessed in different pathophysiological states. The latter issue remains elusive, at least in the context of short-term HRV analysis (up to a few minutes of recordings), due to the difficulty of reliably assessing nonlinearity over short time series, to the proliferation of diverse nonlinear analysis methods each with its own strengths and limitations, and to the changing nature of nonlinear HRV dynamics across states and conditions.
The present study contributes to settle this issue, implementing different of state-of-art nonlinear dynamic measures and comparing them as regards the detection of the presence and the contribution of nonlinear dynamics to short-term HRV. The comparison is performed considering the progression across healthy and pathological states (i.e., aging and myocardial infarction) and investigating the effects on the cardiac dynamics of a specific physiological stressor (i.e., head-up tilt).
\end{quotation}

\section{Introduction}

Human heart rate variability (HRV), commonly assessed measuring the spontaneous beat-to-beat  changes in the duration of the RR interval of the ECG, is the result of the activity of different physiological control systems which operate across multiple time scales to let the body functions adapt to environmental, physical and psychological challenges \cite{berntson1997heart,acharya2006heart}. RR interval fluctuations have been classically represented as a linear superposition of rhythms \cite{akselrod1981power}, leading to remarkable time- and frequency-domain descriptions of the factors contributing to the neuroautonomic modulation of the heart rhythm in healthy conditions, as well as of the alteration of these factors related to a variety of pathological states \cite{baselli1987heart,montano1994power,lombardi1987heart,freeman1991spectral,bigger1992frequency,gujjar2004heart}. Nevertheless, since the cardiac control is typically accomplished through the interaction among multiple complex regulatory mechanisms, including self-sustained oscillators as well as control loops \cite{koepchen1991physiology}, the linear description of the RR interval variability may be severely limited  and disregard significant dynamical features.

As a consequence, a variety of nonlinear approaches to time series analysis have been devised to characterize RR interval fluctuations and extract additional physiological and clinical information from HRV \cite{voss2008methods,stein2005traditional}. A class of these approaches, focused on long-term analyses spanning scales up to several hours, is mainly based on using methods able to assess scaling properties, long-range correlations, and multifractality of the RR time series \cite{karasik2002correlation,Peng1995,ivanov1999multifractality,bernaola2001scale}. These nonlinear methods were often employed with the aim of identifying signatures typical of chaotic dynamics in long-term HRV recordings, leading to an animated discussion of this topic \cite{glass2009introduction}. Besides the presence or absence of chaos \cite{costa1999no,kanters1994lack,govindan1998evidence}, there is substantial consensus about the fact that long-term RR interval time series are nonlinear and multifractal, and that the scaling behavior of HRV is altered with aging or during physical exercise, and under pathological conditions such as myocardial infarction \cite{iyengar1996age,kaplan1991aging,huikuri2000fractal,bernaola2017correlations,gomez2018differences}.

On the other hand, it is also widely accepted that the assessment of HRV over temporal scales ranging from seconds to a few minutes allows the indirect investigation of the mechanisms underlying the short-term cardiovascular control \cite{malik1990heart,CohenTalylor} and this assessment might require nonlinear methods better suited for the evaluation of complex aspects of HRV dynamics. In fact, a number of nonlinear measures have been developed to this end, e.g. based on nonlinear prediction \cite{porta2000prediction,porta2007complexity}, entropy or mutual information \cite{porta2007integrated,xiong2017entropy}, time irreversibility \cite{porta2008temporal,visnovcova2014complexity}, or phase coupling \cite{bai2008nonlinear,chua2008cardiac}. These and other studies have provided ample evidence that changes in nonlinear descriptors of short-term HRV such as complexity or regularity indexes, either induced by the modification of the experimental conditions or determined by spontaneous transitions among physiological states, can be reliably detected and associated to alterations of the autonomic control. Notwithstanding this, the presence and impact of nonlinear dynamics in short-term HRV is still a controversial issue. Some studies suggested that nonlinear components of HRV are of limited importance in resting conditions and are evoked by the presence of a dominant respiratory sinus arrhythmia \cite{porta2000prediction,porta2007complexity}, or in association with respiratory inputs to the cardiovascular system \cite{fortrat1997respiratory,kanters1997influence}. Conversely, other studies assessing temporal asymmetries suggested that nonlinearities are relevant at rest and may be present even in conditions of small respiratory sinus arrhythmia \cite{porta2008temporal}. The contribution of the two branches of the autonomic nervous system to nonlinear HRV dynamics remains elusive and is likely linked to the time scales of their functioning \cite{porta2015complexity}. Moreover, nonlinear dynamics might be sustained by the interaction between sympathetic and parasympathetic activities \cite{bai2008nonlinear}.

Methodologically, it has been suggested that multiple nonlinear components, operating at different scales and possibly interacting with each other, may concur to the generation of short-term HRV \cite{bai2008nonlinear,porta2015complexity}. Since these different components of HRV nonlinear dynamics may be captured in a different way by different metrics, the aim of the present study is to test the hypothesis that distinct types of nonlinear dynamics underlie the HRV dynamics during different physiopatological states. To this end we apply three nonlinear dynamic measures to the RR interval time series measured in young and old healthy subjects, as well as in post acute myocardial infarction (AMI) patients, monitored at rest and during sympathetic activation induced by postural change. Two of the measures quantify the common concepts of time series complexity and regularity, implemented through refined estimation techniques devised recently \cite{porta2019relevance,xiong2017entropy}, while the third measure quantifies the deviation from linearity of the correlation structure of the observed RR series according to a novel Gaussian Linear Contrast method \cite{carpena2019transforming}. The application of these approaches in conjunction with the method of surrogate data \cite{theiler1992testing,schreiber1996improved} allows us to quantify the extent to which nonlinear dynamics impact on short-term HRV in different conditions of autonomic nervous system imbalance, also investigating the effects of age and pathology.
The database used in the study is made publicly available to favor reproducibility and encourage the comparison with different nonlinear dynamic measures.

\section{\label{sec:Nonlinear}Nonlinear Dynamic Measures}
This section describes the methods used in the present work to quantify nonlinear dynamics in the temporal statistical structure of a system evolving in time. Our starting point is an experimental time series $\{s_n\}, n=1,2,\ldots,N$, which is considered as a realization of a stochastic process $S$ describing the evolution over time of an observed dynamical system $\mathcal{S}$.
The process $S$ is considered stationary, so that the random variables obtained sampling the process at the time $n$ (i.e. $S_n,n=1,2,\ldots,N$), are identically distributed with marginal distribution of probability density $f(\cdot)$ and cumulative distribution $F(\cdot)$. Moreover, without loss of generality, we assume that each $S_n$ has zero mean and unit standard deviation.

To assess nonlinear dynamics in the stochastic process $S$ we look at its temporal correlation structure: while for purely linear dynamics the dependence between $s_n$ and $s_{n-\ell}$ is linear for any lag $\ell$, in the case of nonlinear dynamics such dependence cannot be studied only in terms of linear correlations. In the first two methods considered, nonlinear correlations are investigated within an information-theoretic framework, separating the present state of the system $S_n$ from its past states $S_n^{-}=[S_1, \ldots, S_{n-2}, S_{n-1}]$ and quantifying their information content in terms of entropy measures \cite{xiong2017entropy}. In fact, when the system transits from past states to a new state, new information is produced in addition to the information that is already carried by the past states. The rate of generation of new information is inversely related to the strength of nonlinear correlations in the process, while the information shared between the present and the past variables is directly related to such correlations. On this basis, the measures of conditional entropy (Sect. \ref{sec:NCI}) and information storage (Sect. \ref{sec:IS}) assess nonlinear correlations quantifying respectively the new information contained in $S_n$ but not in $S_n^{-}$, and the amount of information carried by $S_n$ that can be explained by $S_n^{-}$.

The third method takes its roots on the observation that a purely linear stochastic process is considered to have: (i) Gaussian marginal distribution; and (ii) only linear correlations. Therefore, if $S$ is a nonlinear process it should lack one or both of these properties, so that the marginal distribution of $S_n$ is not Gaussian and/or the correlations are not linear. However, these two properties are not equally important: many authors consider \cite{theiler1992testing, schreiber1996improved} that a non Gaussian distribution is simply due to a static transformation in the output of the dynamical system (for example by a filter) and that the possible nonlinearity is only due to the nature of the correlations, which truly reflect the dynamics. This is the approach we adopt in the Gaussian Linear Contrast method (GLC) presented in Sect. \ref{sec:GLC}.

\subsection{\label{sec:NCI}Complexity Index based on Local Sample Entropy}
The information-theoretic assessment of nonlinear correlations in a dynamic process is based on applying the concepts of entropy and conditional entropy to the random variables representing the present and past states of the process. Given two generic continuous (possibly vector) random variables $X$ and $Y$, the entropy of $X$ and the conditional entropy of $X$ given $Y$ are defined as
\begin{equation}
H(X)=-\mathbb{E}[\log f(x)]=-\int_{\mathcal{D}_X} f(x)\log f(x)dx,
\label{entropy}
\end{equation}
\begin{equation}
H(X|Y)=-\mathbb{E}[\log f(x|y)]=H(X,Y)-H(Y),
\label{conditional_entropy}
\end{equation}
where $\mathcal{D}_X$ is the domain of $X$, $f(x)$ and $f(x|y)$ are the probability density of $X$ and the conditional probability of $X$ given $Y$, and $\mathbb{E}[\cdot]$ is the expectation operator; the term $H(X,Y)$ in (\ref{conditional_entropy})  is the joint entropy of $X$ and $Y$, obtained generalizing (\ref{entropy}) to the joint probability density $f(x,y)=f(x|y)f(y)$. Particularizing these definitions to the variables $S_n$ and $S_n^{-}$ describing the present and the past states of the process $S$, the conditional entropy becomes 
\begin{equation}
H(S_n|S_n^{-})=H(S_n^{-},S_n)-H(S_n^{-})
\label{eq:CE}
\end{equation}
The conditional entropy quantifies the amount of information contained in the present of the process that cannot be explained by its past history: if the process is fully random, the system produces information at the maximum rate, yielding maximum conditional entropy; if, on the contrary, the process is fully predictable, the system does not produce new information and the conditional entropy is zero.

In the present work, practical computation of the conditional entropy is performed adopting kernel estimates of the probability density functions \cite{xiong2017entropy}. In particular, we make use of the well known Sample Entropy index \cite{richman2000physiological}, improved through the implementation of a local version of the estimator \cite{porta2019relevance}. The Sample Entropy (SampEn) estimates $H(S_n|S_n^{-})$ in (\ref{eq:CE}) first truncating $S_n^{-}$ to  $S_n^{m}=[S_{n-1},S_{n-2},\ldots,S_{n-m}]$, and then approximating
$H(S_n^{m},S_n)$ and $H(S_n^{m})$ as the negative logarithm of the average joint probability of finding a pattern in the neighborhood of the reference pattern with a tolerance \textit{r} in the $(m+1)$dimensional and $m-$dimensional embedding space, namely 
\begin{equation}
{\rm SampEn}(m,r,n)=- \ln \left\langle p(S_n^{m},S_n) \right\rangle + \ln \left\langle p(S_n^{m}) \right\rangle
\label{eq:SampEn},
\end{equation}
where $p(S_n^{m},S_n)$ is the probability that the pattern $S_n^{m} \oplus S_n$ assumes the value
$s_n^{m} \oplus s_n$ , $p(S_n^{m})$  is the probability that the pattern $S_n^{m}$ takes the value $s_n^{m}$ 
and $\left\langle \cdot \right\rangle$ performs the average over time 
(i.e., over all values $s_n^{m} \oplus s_n$). 
SampEn is a robust estimator of irregularity given that the log-of-zero situation is extremely unlikely because the logarithm is applied to the average of a quantity that has 0 as the lowest bound. However, as a consequence of computing an average over time, SampEn has the disadvantage to be a global marker of irregularity that might not represent reliably the local behavior in the neighborhood of a specific pattern and blur nonlinear features \cite{porta2019relevance}. A local version of SampEn (LSampEn) was proposed in Ref. \cite{porta2019relevance} by directly approximating $H(S_n|S_n^{m})$ instead of its constituents (i.e., $H(S_n^{m},S_n)$ and $H(S_n^{m})$) as
\begin{equation}
{\rm LSampEn}(m,r,n)=- \ln \left\langle p(S_n|S_n^{m}) \right\rangle
\label{eq:LSampEn},
\end{equation}
where $p(S_n|S_n^{m})$ is the conditional probability that the current state $S_n$ assumed the value $s_n$ given that the past state $S_n^{m}$ is $s_n^{m}$. The average operator makes the estimator robust against the log-of-zero situation and the estimation of $p(S_n|S_n^{m})$ renders LSampEn a local estimator of irregularity given that the quantity being averaged referred specifically to the reference pattern $S_n^{m}$. To limit the consequence that, when solely $S_n^{m}$ is found in the neighborhood of $S_n^{m}$,  $p(S_n|S_n^{m})$ is unreliably high \cite{porta1998measuring}, we applied the correction proposed by Porta et al. \cite{porta2019relevance}, namely in this unfortunate case $p(S_n|S_n^{m})$ is set to $(N-m+1)^{-1}$ corresponding to the maximum uncertainty computable over the series. The resulting estimator, applied to the time series reduced to unit variance, is denoted as Normalized Complexity Index (NCI) \cite{porta2019relevance}.

\subsection{\label{sec:IS}Regularity Index based on Information Storage}
Information measures can be exploited also for evaluating in a direct way the strength of nonlinear correlations in the dynamical structure of a stochastic process, so that to assess its degree of regularity. To this end, a relevant entropy measure is the so-called information storage, which quantifies the amount of information shared between the present and the past observations of the considered process. The information storage of the process $S$ is defined as:
\begin{eqnarray}
I(S_n;S_n^{-})=H(S_n)+H(S_n^{-})-H(S_n^{-},S_n),
\label{eq:IS}
\end{eqnarray}
where $I(\cdot;\cdot)$ denotes mutual information.
The information storage reflects the degree to which information is preserved in a time-evolving system \cite{wibral2014local}. As such, it measures how much of the uncertainty about the present can be resolved by knowing the past: if the process is fully random, the past gives no knowledge about the present, so that the information storage is zero; if, on the contrary, the process is fully predictable, the present can be fully predicted from the past, which results in maximum information storage. Note that information storage and conditional entropy of a dynamic process are inversely related to each other, and depend on the entropy of the present state of the process through the equation
$I(S_n;S_n^{-})+H(S_n|S_n^{-})=H(S_n)$.

In practical analysis, the information storage can be estimated from a time series of finite length following the same principles of conditional entropy estimation. These include the use of a finite number of samples in the past to approximate the history of the observed process (i.e., $S_n^{-}$ is truncated to $S_n^{m}=[S_{n-1},S_{n-2},\ldots,S_{n-m}]$),
and the adoption of non-parametric estimators of the probability density functions involved in the computation of $I(S_n;S_n^{m})$. However, since computation of the measure defined in (\ref{eq:IS}) requires to estimate three entropy terms involving variables of different dimensions, and since the bias of entropy estimates depends strongly on the dimension, implementation of standard histogram or kernel-based methods typically results in inaccurate estimates of the information storage \cite{faes2015estimating, xiong2017entropy}. Here, to overcome this limitation, we resort to nearest neighbor entropy estimation \cite{kozachenko1987sample} and implement a strategy for bias compensation specific of mutual information estimates \cite{kraskov2004estimating}. The nearest neighbor entropy estimate of a generic $d$-dimensional random variable $X$ can be obtained from a set of realizations $\{x_1,x_2,\ldots,x_N\}$ of the variable as \cite{kozachenko1987sample} 
\begin{equation}
H(X)=\psi(N)-\psi(k)+d\langle\ln\varepsilon_n\rangle,
\label{eq:H_knn}
\end{equation}
where $\psi$ is the digamma function, $\varepsilon_n$ is twice the distance between the outcome $x_n$ and its $k^{\rm th}$ nearest neighbor computed according to the maximum norm (i.e., taking the maximum distance of the scalar components), and $\langle \cdot \rangle $ stands for average over $N$ outcomes. Then, the information storage could be computed applying (\ref{eq:H_knn}) to the three terms in (\ref{eq:IS}). However, doing so would result in different distance lengths when approximating the probability density in different dimensions, and this would introduce different estimation biases that cannot be compensated by taking the entropy differences.
To keep the same distance length in all explored spaces, we perform a neighbor search only in the highest-dimensional space (the one spanned by the realizations of $S_n^{m},S_n$) and then project the distances found in this space to the lower-dimensional spaces (those spanned by the realizations of $S_n^{m}$ and $S_n$), keeping these distances as the range within which neighbors are counted.
Specifically, the knn estimate of $H(S_n^{m},S_n)$ is computed through the neighbor search:
\begin{equation}
H(S_n,S_n^m)=\psi(N)-\psi(k)+(m+1)\langle\ln\varepsilon_n\rangle,
\label{Eq:XnXnm_knn}
\end{equation}
where $\varepsilon_n$ is twice the distance from $(S_n,S_n^m)$ to its $k^{\rm th}$ nearest neighbor, and then, given the distances $\varepsilon_n$, the entropies in the lower-dimensional spaces are estimated through a range search:
\begin{equation}
H(S_n^m)=\psi(N)-\psi(N_{S_n^m})+m\langle\ln\varepsilon_n\rangle,
\label{Eq:Xm_knn}
\end{equation}
\begin{equation}
H(S_n)=\psi(N)-\psi(N_{S_n})+\langle\ln\varepsilon_n\rangle,
\label{Eq:Xn_knn_projected}
\end{equation}
where $N_{S_n}$ and $N_{S_n^m}$ are the number of points whose distance from $S_n$ and $S_n^m$, respectively, is smaller than $\varepsilon_n/2$.
Finally, our estimate of the information storage is obtained subtracting Eq.~(\ref{Eq:XnXnm_knn}) from the sum of
Eqs.~(\ref{Eq:Xm_knn}) and ~(\ref{Eq:Xn_knn_projected}) \cite{porta2015disentangling}:
\begin{equation}
IS=\psi(N)+\psi(k)-\langle\psi(N_{S_n^m})\rangle-\langle\psi(N_{S_n})\rangle.
\label{Eq:information_storage_knn}
\end{equation}

\subsection{\label{sec:GLC}Nonlinearity Index based on Gaussian Linear Contrast}

As we stated above, GLC assesses nonlinearities related only to the nature of the correlations and not to the non-Gaussianity of the data.
Let us consider an experimental time series $\{s_n\}$  ($n=1,2,\ldots,N$), with non-Gaussian marginal distribution. The observed autocorrelation
funtion of $\{s_n\}$ is given by
\begin{equation}
C_{\rm obs}(\ell)=\langle s_n s_{n+\ell} \rangle
\end{equation}
Using $C_{\rm obs}(\ell)$, GLC tries to determine if $\{s_n\}$ is 
originated from a Gaussian time
series $\{z_{G,n}\}$ ($n=1,2,\ldots,N$) with only \textsl{linear 
correlations}, which have been transformed to have the observed marginal 
distribution
of the experimental time series. If this is the case, then GLC assumes 
that $\{s_n\}$ is linear, and is non-linear otherwise.

The theoretical background of the GLC method is the following. Let us 
consider a pair of correlated Gaussian variables $x_G$ and $y_G$,  both 
of ${\cal{N}}(0,1)$ type, so that their corresponding probability 
density and cumulative distribution are the standard Gaussian 
$\varphi(x_G)$ and $\Phi(x_G)$. We assume that  $x_G$ and $y_G$ are only 
linearly correlated, with a correlation value $C_G$, i.e.
\begin{equation}
C_G=\langle x_G y_G \rangle \label{gcor}
\end{equation}
This is equivalent to affirm that the joint distribution of $x_G$ and 
$y_G$ is the bivariate Gaussian distribution $\varphi_2(x_G,y_G,C_G)$.
Then, we transform $x_G$ and $y_G$ to the variables $x$ and $y$, which 
follow the marginal distribution of the experimental time series. This 
can be done
with the usual method:
\begin{equation}
x=F^{-1}[\Phi(x_G)], \,\,\, y=F^{-1}[\Phi(y_G)] \label{tr}
\end{equation}
with $F^{-1}(\cdot)$ the inverse cumulative distribution of the 
experimental time series. Since $F^{-1}$
is fixed by $\{s_i\}$, the linear correlation $C$ between $x$ and $y$, 
i.e. $C=\langle x y \rangle$
depend solely on the $C_G$ value. Indeed, since $x$ and $y$ depend 
formally on $x_G$ and $y_G$ (Eq. (\ref{tr})) with joint distribution 
$\varphi_2(x_G,y_G,C_G)$,
$C$ can be calculated as \cite{carpena2019transforming}
\begin{widetext}
\begin{equation}
C(C_G)\equiv\langle x y \rangle 
=\int_{-\infty}^{\infty}\int_{-\infty}^{\infty} F^{-1}(\Phi(x_G))\, 
F^{-1}(\Phi(y_G))\, \varphi_2(x_G,y_G,C_G)\, dy_G \, dx_G \label{xy}
\end{equation}
\end{widetext}
Solving numerically the previous integral for a dense set of $C_G$ 
values in the interval $(-1,1)$ we characterize the $C(C_G)$ function, 
which contains the information on
how the linear Gaussian correlations are transformed when the 
distribution of the variables is transformed from Gaussian to the 
experimental distribution.

These results can be extrapolated straightforwardly to time series. Let 
us consider a Gaussian, linearly correlated time series $\{z_{G,n}\}$, 
with autocorrelation function $C_G(\ell)$ given by
$C_G(\ell)\equiv \langle z_{G,n} z_{G,n+\ell}\rangle$. Note that 
$z_{G,n}$ and $z_{G,n+\ell}$ are equivalent to $x_G$ and $y_G$ in Eq. 
(\ref{gcor}). Then, let us transform
$\{z_{G,n}\}$ into a time series $\{z_n\}$ with the same marginal 
distribution of the experimental time series using Eq. (\ref{tr}) for 
each $z_{G,n}$ value.
The autocorrelation function $C(\ell)$ of $\{z_n\}$ can be then 
calculated using Eq. (\ref{xy}) simply by replacing $x_G$, $y_G$ and 
$C_G$ by $z_{G,n}$, $z_{G,n+\ell}$ and $C_G(\ell)$
respectively. In other words, once the $C(C_G)$ function is known by 
using Eq. (\ref{xy}) (which only requires the marginal distribution of 
the experimental time series), then $C(\ell)=C(C_G(\ell))$. This last 
equality holds if, and only if, the non-Gaussian time series $\{z_n\}$ 
really comes via the transformation (\ref{tr}) from the Gaussian and 
linearly correlated series $\{z_{G,n}\}$ since this is the condition 
used in Eq. (\ref{xy}) to determine $C(C_G)$. This property is the key 
point in the GLC
method.

With this theoretical background, the steps to apply the GLC method on 
an experimental time series $\{s_i\}$ are the following:
\begin{itemize}
     \item[(i)] Determine the observed autocorrelation function $C_{\rm 
obs}(\ell)$ of the experimental time series $\{s_n\}$.
     \item[(ii)] Transform $\{s_n\}$ to have Gaussian distribution using 
the inverse of the transformation in Eq.(\ref{tr}), and calculate its 
autocorrelation function $C_{G'}(\ell)$. Note that if $\{s_n\}$ has been 
obtained from a Gaussian time series using the transformation 
(\ref{tr}), symply by inverting the transformation the hypothetical 
original Gaussian time series is recovered.  The knowledge of $C_{\rm 
obs}(\ell)$ and $C_{G'}(\ell)$ for each $\ell$ allows to obtain the 
function $C_{\rm obs}(C_{G'})$.

     \item [(iii)] Obtain the real $C(C_{G})$ function using 
Eq.(\ref{xy}) by giving to $C_G$ a great number of values in the 
interval $(-1,1)$. In practice, and specially for short experimental 
time series, the numerical solution of the integral might be a harsh 
task: due to finite size effects it can be difficult to correctly 
estimate $F^{-1}$. Then, to calculate $C(C_{G})$ we adopt a different 
strategy: we use autoregressive processes of order 1 (AR1) with the same 
size as $\{s_n\}$). An AR1 process is defined as:
$z_{G,n}=\varphi z_{G,n-1} + \eta_{n}$ where $\{\eta_n\}$ is a Gaussian 
${\cal{N}} (0,1)$ white noise and $\varphi \in(-1,1)$ is a constant. AR1 
processes are Gaussian with purely linear correlations,
     with generic autocorrelation $C_{G}(\ell)=\varphi^{\ell}$, so that 
changing the $\varphi$ value we can obtain any value of Gaussian 
correlation in the interval (-1,1). Thus, we generate a large set of
     AR1 processes for different $\varphi$ values, and calculate the 
autocorrelation function of all of them obtaining a huge amount of data 
points densely populating the $(-1,1)$ Gaussian correlation interval. 
Then, we transform all AR1 processes using Eq.(\ref{tr}) to time series 
having the marginal distribution of $\{s_n\}$, and also calculate the 
autocorrelation function $C(\ell)$ for all series. Note that each 
$C(\ell)$ value is the image of a Gaussian autocorrelation value. 
Finally, we bin the Gaussian correlation interval $(-1,1)$ into $0.01$ 
length bins, and put in each one the images of all the Gaussian 
correlation values contained in the bin. The average of all the images 
in the respective bin gives the $C$ value corresponding to the Gaussian 
correlation at the center of the bin, so that finally we have a numeric 
determination of the $C(C_G)$ function.

     \item[(iv)] If the experimental time series $\{s_n\}$ is really 
obtained by transforming a Gaussian time series, then the Gaussian 
series is the one determined in step (ii), with autocorrelation function 
$C_{G'}(\ell)$, and the observed autocorrelation are given by $C_{\rm 
obs}(\ell)=C_{\rm obs}(C_{G'}(\ell))$. However, the expected 
correlations in $\{s_n\}$ \textsl{if the Gaussian series is purely 
linear}, $C_{\rm lin} (\ell)$, should be given by the $C(C_G)$ function 
determined in step (iii) evaluated at the $C_{G'}(\ell)$ values, i.e. 
$C_{\rm lin}(\ell)=C(C_{G'}(\ell))$. The series
     $\{s_n\}$ is linear when $C_{\rm obs}(\ell)=C_{\rm lin}(\ell)$ and 
is not linear otherwise. In this way, to quantify the nonlinearity of  
$\{s_n\}$ we can define the GLC non-linearity index as
\begin{equation}
GLC\equiv\sum_{\ell=1}^{\ell_{\rm m}}|C_{\rm obs}(\ell)-C_{\rm lin}(\ell)|
\end{equation}


\end{itemize}

\section{\label{sec:DetQuNL}Detection and Quantification of Nonlinearity}

The existence of nonlinear dynamics in the considered time series was investigated in accordance with the method of surrogate data \cite{theiler1992testing}. This approach is based on: (i) a null hypothesis to be rejected; (ii) a surrogate data set constructed in accordance with the null hypothesis; (iii) a discriminating statistic that has to be calculated on original and surrogate series; and (iv) a statistical test allowing to reject or confirm the null hypothesis.

The null hypothesis set in our case is that the investigated time series is a realization of a Gaussian stochastic process (fully described by linear temporal autocorrelations), eventually measured through a static and possibly nonlinear transformation distorting the Gaussian distribution.

The surrogate time series were generated in order to preserve the linear autocorrelation structure as well as the marginal distribution of the original time series. This was achieved through the iteratively refined amplitude adjusted Fourier Transform (IAAFT) method \cite{schreiber1996improved}. The method is an improvement of the Fourier transform (FT) method \cite{theiler1992testing}, which generates surrogate time series by computing the FT of the original series, substituting the Fourier phases with random numbers uniformly distributed between $0$ and $2\pi$, and finally performing the inverse FT. Since the FT method distorts the amplitude distribution of the original process when such a distribution is not Gaussian, the IAAFT method is followed implementing an iterative procedure that alternatively constrains the surrogate series to have the same power spectrum (by replacing the squared Fourier amplitudes of the candidate surrogate series with those of the original series) and to have the same amplitude distribution (by a rank ordering procedure) of the original series.

As discriminating statistic we employ each of the three nonlinear indexes presented in Sect. (\ref{sec:Nonlinear}), i.e., the nornalized complexity index (NCI) based on local Sample Entropy, the regularity index based on information storage (IS), and the nonlinear index based on Gaussian Linear Contrast (GLC).

As statistical test, we perform a nonparametric test based on percentiles. The test compares the selected nonlinear index, here denoted generically as $NI$, when calculated on the original time series ($NI_o$) and when calculated on $n_s$ surrogate time series ($n_s=100$ in this work) generated under the null hypothesis.
Specifically, $NI_o$ was compared with a threshold for significance $NI_{\alpha}$ extracted from the empirical distribution of $NI$ over the surrogates setting a prescribed confidence level $\alpha$ ($\alpha=0.05$ in this work).
In the case of the NCI index measuring the complexity of a time series, the index is expected to decrease in the presence of nonlinear dynamics compared to linear time series; therefore, $NI_{\alpha}$ was set at the $100\cdot\alpha$-percentile of the distribution of $NI$ over the surrogates and the null hypothesis was rejected if $NI_o<NI_{\alpha}$.
In the case of the IS and GLC indexes measuring the regularity of a time series or the amount of nonlinear correlations, the indexes are expected to increase in the presence of nonlinear dynamics compared to linear time series; therefore, $NI_{\alpha}$ was set at the $100\cdot(1-\alpha)$-percentile of the distribution of $NI$ over the surrogates and the null hypothesis was rejected if $NI_o>NI_{\alpha}$.

Besides detecting the presence of nonlinearity, the analysis of original and surrogate time series was exploited also to quantify the `extent' of nonlinearity in the investigated time series. This was performed comparing the index $NI_o$ computed on the original, possibly nonlinear time series, with the median $NI_m$ of its values computed on the set of surrogate time series. The difference with the median, defined as $\Delta NI=NI_m-NI_o$ in the case of the complexity index (i.e., when $NI=NCI$), and defined as $\Delta NI=NI_o-NI_m$ in the case of the two other indexes (i.e., when $NI=IS$ or when $NI=GLC$), was taken as a measure of the amount of nonlinearity in the observed time series.

\section{Patients, Experimental Protocol and Data Analysis}

The time series analyzed in this study belong to a database to analyze the effects of aging and myocardial infarction on cardiovascular interactions \cite{nollo2002evidence}. The database consists on heart rate variability measured in a group of 35 post-acute myocardial infarction (\textit{AMI}, $58.5 \pm 10.2$ years old) patients examined about 10 days after AMI, and in two control groups formed by 12 age-matched old healthy subjects (\textit{Old}, $63.1 \pm 8.3$ years) and by 19 young healthy subjects (\textit{Young}, $25.0 \pm 2.6$ years). Eight out of 35 post-AMI patients were initially under beta-locker therapy, but they discontinued the treatment two half-lives before the recording session. Control subjects were normotensive and free from any known disease based on anamnesis and physical examination at the time of the study.

After a period of 15 min for subject stabilization, the electrocardiogram (lead II ECG) was recorded for 10 min in the supine rest position, followed by 10 min of passive head-up tilt at $60$ degrees. All ECG signals were digitized with a 12 bit resolution and 1-KHz sampling rate. After detecting the QRS complex on the ECG and locating the R apex through template matching, heart period variability was measured on a beat-to-beat basis calculating the sequence of the time intervals occurring between pairs of consecutive R peaks (RR intervals). The series were then cleaned up from artifacts, windowed to $N=300$ points for each condition (\textit{rest}, \textit{tilt}), and detrended by a high-pass filter to fulfill 
stationarity criteria \cite{xiong2017entropy, nollo2000synchronization}. Normalized time series were eventually obtained by subtracting the mean values and dividing by the standard deviation.

For each subject and condition, analysis of nonlinearity was performed using the three methods described in Sect. \ref{sec:Nonlinear} and performing the tests described in Sect. \ref{sec:DetQuNL}. NCI and IS indexes were computed using standard values for the free parameters of entropy estimators applied to short time series \cite{richman2000physiological,pincus1994physiological}, namely using $m=2$ values to approximate the past history of the process, setting a tolerance $r=0.2\sigma$ to define similarity in Sample Entropy analysis (where $\sigma$ is the standard deviation of the series equal to 1 after normalization), and employing $k=10$ neighbors in the distance-based entropy estimations. Distances between patterns were obtained using the Eucilidean norm in the kernel estimator used to compute NCI \cite{porta2019relevance}, and the maximum norm in the nearest-neighbor estimator used in IS \cite{xiong2017entropy}.
In the computation of the GLC index, taking into account the short size of the time series ($N=300$) and to align with the other measures, we choose $\ell_{\rm max}=m=2$ to limit spurious results induced by the fact that the autocorrelation function tends to reach quickly the noise level.

For each assigned index (NCI, IS, GLC), the statistical significance of its changes across groups (\textit{Young},\textit{Old},\textit{AMI}) was assessed by the Kruskal Wallis test followed by the Wilcoxon rank sum test (Mann-Whitney U test) to detect pairwise differences (\textit{Young} vs. \textit{Old}, \textit{Young} vs. \textit{AMI}, or \textit{Old} vs. \textit{AMI}).
For an assigned index and group, the statistical significance of the differences between conditions (\textit{rest} vs. \textit{tilt}) was assessed by the paired Wilcoxon signed rank test.
We computed also the percentage of subjects belonging to each group for which the null hypothesis of linear Gaussian dynamics was rejected in the two conditions; then, statistically significant variations between two groups in a given condition were assessed using the chi-square test for proportions, while significant variations  between conditions for a given group were assessed using the McNemar test for paired proportions.

\section{Results}

Figure 1 reports an illustrative example of the application of the three considered nonlinear dynamic measures to the RR interval time series obtained in the two analyzed conditions (\textit{rest}, \textit{tilt}) on representative subjects belonging to the three considered groups (\textit{Young}, \textit{Old}, \textit{AMI}).
Considering the two entropy measures, opposite response to the change in condition are observed consistently for the three cases, with lower values of NCI and higher values of IS measured during \textit{tilt} compared to \textit{rest}.
On the contrary, moving from \textit{rest} to  \textit{tilt} the nonlinear dynamic measure based on GLC decreases slightly for the  \textit{Young} subject (circles), decreased more consistently for the \textit{Old} subject (squares), and increases for the \textit{AMI} patient (triangles).
Moreover, the comparison between the original value of a measure and its distribution on the surrogate time series reveals the different ability to detect nonlinear dynamics of the different measures. In particular, in both the experimental conditions nonlinear dynamics are detected only by the Information storage in the \textit{Young} subject (Fig. 1a) and only by the Gaussian Linear Contrast method in the \textit{AMI} patient (Fig. 1c), while all measures detect the presence of nonlinear dynamics in the \textit{Old} subject (Fig. 1b, NCI and IS in both conditions and GLC only at \textit{rest}).

\begin{figure*}[ht]
\includegraphics[width=15cm]{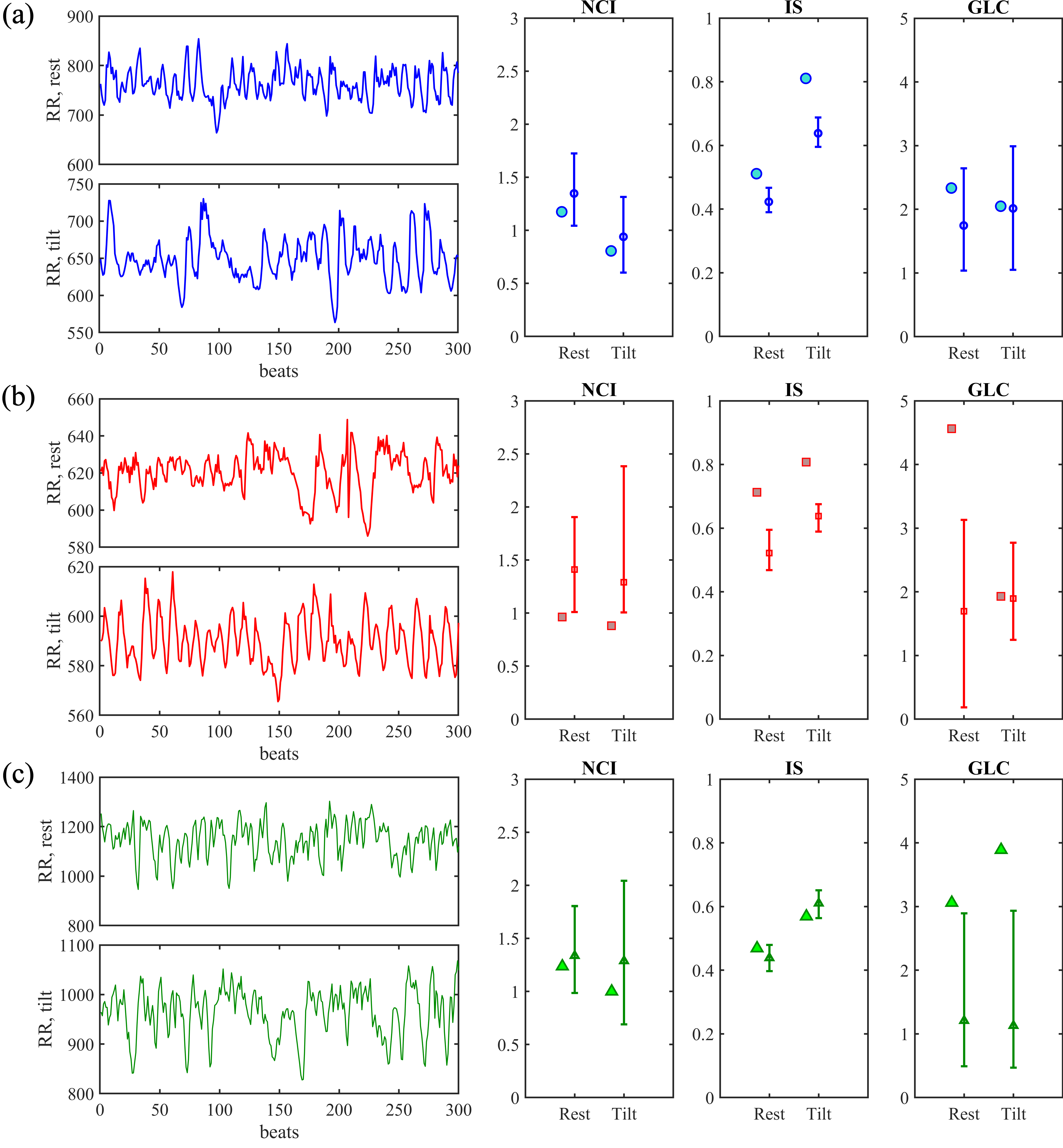}
\caption{\label{fig:example} Computation of nonlinear dynamic measures on heart period time series measured for a representative young subject (a, blue), old subject (b, red) and post AMI patient (c,green). For each subject, the time series of the RR interval measured in the two experimental conditions are reported on the left (above: \textit{rest}; below: \textit{tilt}), and the values of the nonlinear dynamic measures obtained with the three considered methods (NCI: Normalized Complexity Index; IS: Information Storage; GLC: Gaussian Linear Contrast) are reported on the right (markers: original values; error bars: $5^{th},50^{th},95^{th}$ percentiles of the  distribution over 100 surrogates).}
\end{figure*}

Most of the trends observed for the representative subjects described above are reflected at the population level, as reported in Figure 2 showing the distributions across subjects and conditions of the three nonlinear dynamic measures. The indexes based on conditional entropy and mutual information display opposite trends in response to the change of posture: the transition from \textit{rest} to  \textit{tilt} is associated with a statistically significant decrease of the complexity index (NCI, Fig. 2a) and a statistically significant increase of the information storage (IS, Fig. 2b) in both \textit{Young} and \textit{AMI} groups, while no significant changes are detected for both measures in the \textit{Old} group. Moreover, during \textit{tilt} NCI is significantly higher, and IS is significantly lower, in \textit{Old} and \textit{AMI} compared to \textit{Young} (Fig. 2a,b). As to the GLC measure, it changes with the experimental condition in different ways for the different groups (Fig 2c): moving from \textit{rest} to  \textit{tilt} the measure decreases significantly in the \textit{Young} subjects, does not change significantly in the \textit{Old} subjects, and increases significantly in the \textit{AMI} patients. The increase displayed with \textit{tilt} is such that the GLC measure becomes significantly higher in \textit{AMI} compared to \textit{Young}. 

Fig. 3 depicts the results of the analysis performed considering the deviation of each nonlinear dynamic measure from its median level assessed on linear Gaussian surrogates. We find that measures based on conditional entropy and mutual information decrease significantly, in \textit{Young} healthy subjects, with the transition from \textit{rest} to  \textit{tilt}, while no significant changes are observed for \textit{Old} subjects and \textit{AMI} patients (Fig. 3a,b). Moreover,
\textit{AMI} patients display significantly lower IS values compared to \textit{Young} subjects (Fig. 3b). On the other hand, the deviation from the median surrogate value of the GLC measure exhibits similar variations to those reported in Fig. 2c for the original values, as the index decreases significantly from \textit{rest} to  \textit{tilt} in the \textit{Young} subjects, and is significantly higher for \textit{AMI} patients than for \textit{Young} subjects (Fig. 3c).

\begin{figure}[ht]
\includegraphics[clip,width=0.65\columnwidth]{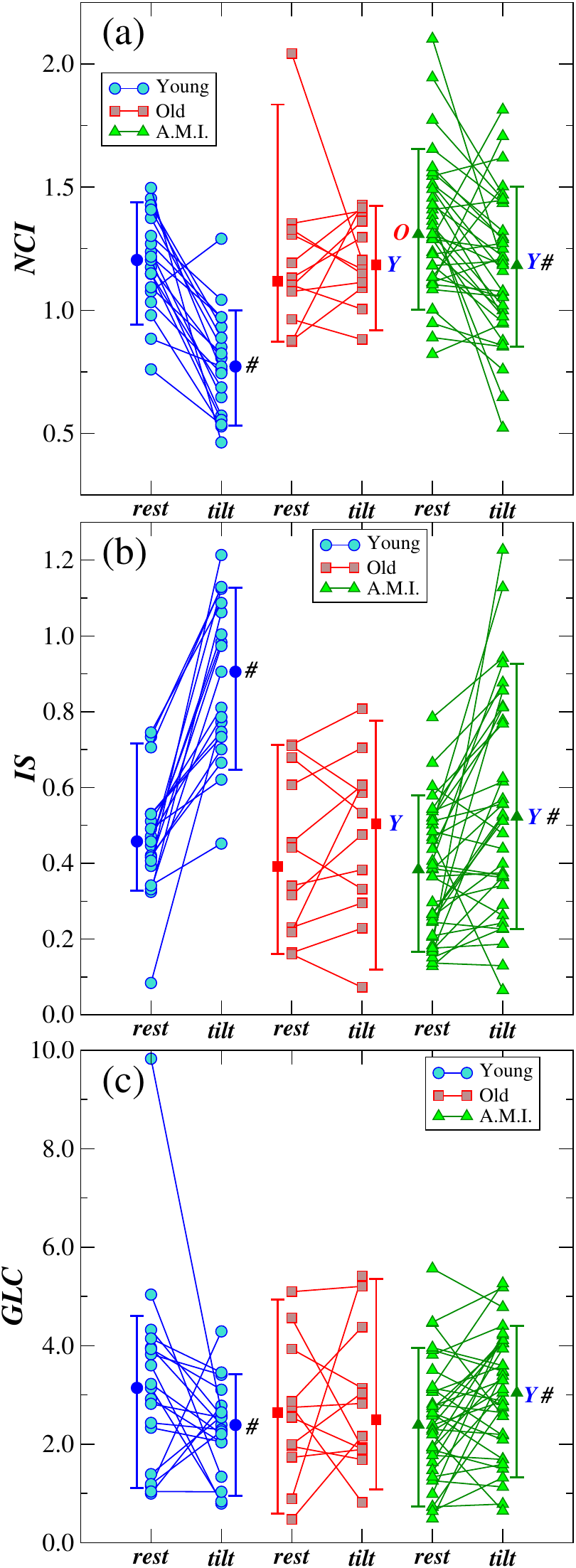}
\caption{\label{fig:index} Distribution of the nonlinear dynamic measures computed with the three proposed methods (a, Normalized Complexity Index; b, Information Storage; c, Gaussian Linear Contrast) over the heart period time series of Young subjects (blue circles), Old subjects (red squares) and post AMI patients (green triangles) in the $rest$ and $tilt$ conditions. Distributions are shown indicating both individual values (markers) and $5^{th},50^{th},95^{th}$ percentiles of the distributions across subjects (error bars). Statistical analysis: $\#, p<0.05 rest $ vs. $tilt$; $Y, p<0.05$ vs. Young; $O, p<0.05$ vs. Old.}
\end{figure}

\begin{figure}[ht]
\includegraphics[clip,width=0.65\columnwidth]{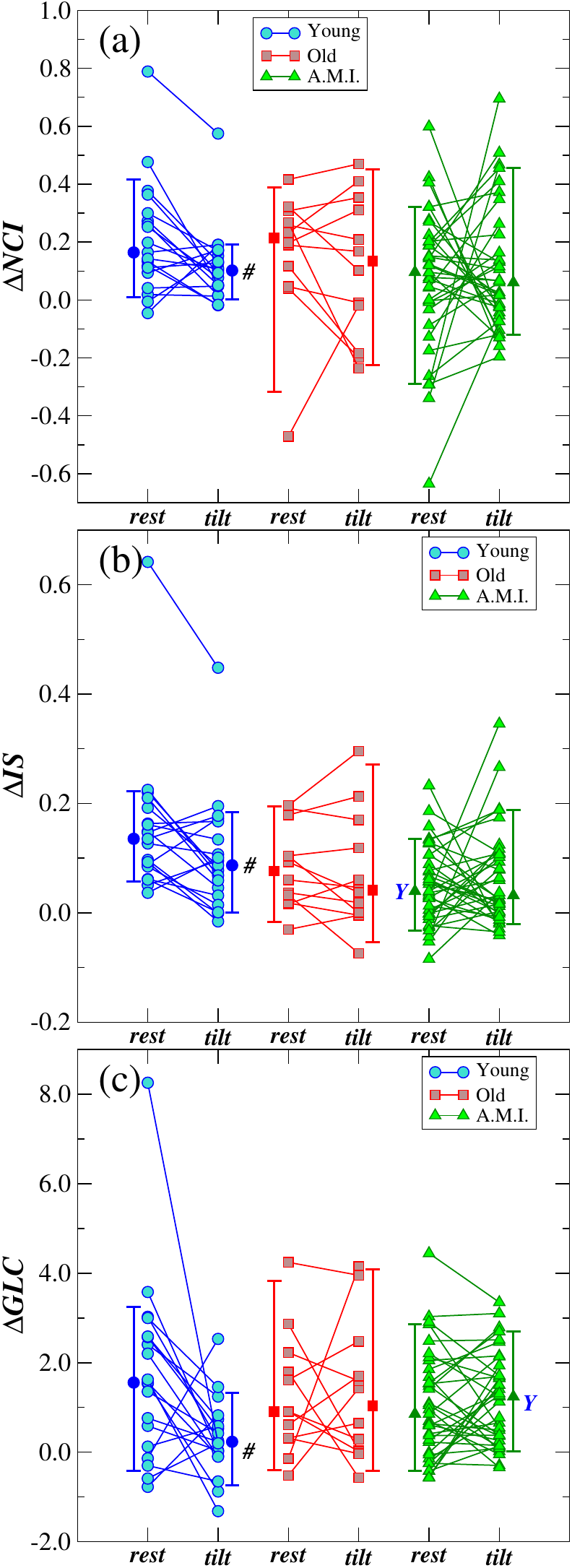}
\caption{\label{fig:deltaindex} Distribution of the measures quantifying the extent of nonlinearity (difference with the median) computed with the three proposed methods (a, Normalized Complexity Index; b, Information Storage; c, Gaussian Linear Contrast) over the heart period time series of Young subjects (blue circles), Old subjects (red squares) and post AMI patients (green triangles) in the $rest$ and $tilt$ conditions. Distributions are shown indicating both individual values (markers) and $5^{th},50^{th},95^{th}$ percentiles of the distributions across subjects (error bars). Statistical analysis: $\#, p<0.05 rest$ vs. $tilt$; $Y, p<0.05$ vs. Young.}
\end{figure}

Fig. 4 reports the the relevance of nonlinear dynamics in each group and experimental condition, measured as the percentage of subjects for which the value of the considered nonlinear dynamic measure computed for the original RR series is deemed (with 5 \% significance) as not drawn from the distribution of the index derived from the surrogate RR series. The conditional entropy measure is associated with nonlinear dynamics in less than half of the subjects in each group, as the NCI index is found below the $5^{th}$ percentile of its surrogate distribution in $\sim 35 \%$ of \textit{Young} subjects, $\sim 45 \%$ of \textit{Old} subjects, and $\sim 25 \%$ of \textit{AMI} patients (with no substantial differences between conditions, Fig. 4a).
The mutual information measure detects a considerably higher percentage of nonlinear dynamics, as the IS index is found above the $95^{th}$ percentile of its surrogate distribution in more than half of the subjects in all groups and conditions (Fig. 3b). In the \textit{Young} group, the IS index is larger than the significance threshold in $95 \%$ of the subjects at rest and in $\sim 70 \%$ of the subjects during tilt; in the \textit{Old} and \textit{AMI} groups the index is significantly lower during both conditions (Fig. 4b).
The Gaussian Linear contrast approach detects nonlinear dynamics in $\approx 30-60 \%$ of subjects in all groups and conditions (Fig. 4c). Moving from \textit{rest} to \textit{tilt}, the number of subjects with nonlinear dynamics detected by the GLC measure decreases in \textit{Young}, while it increases in \textit{Old} and \textit{AMI}.

\begin{figure}[ht]
\includegraphics[clip,width=0.65\columnwidth]{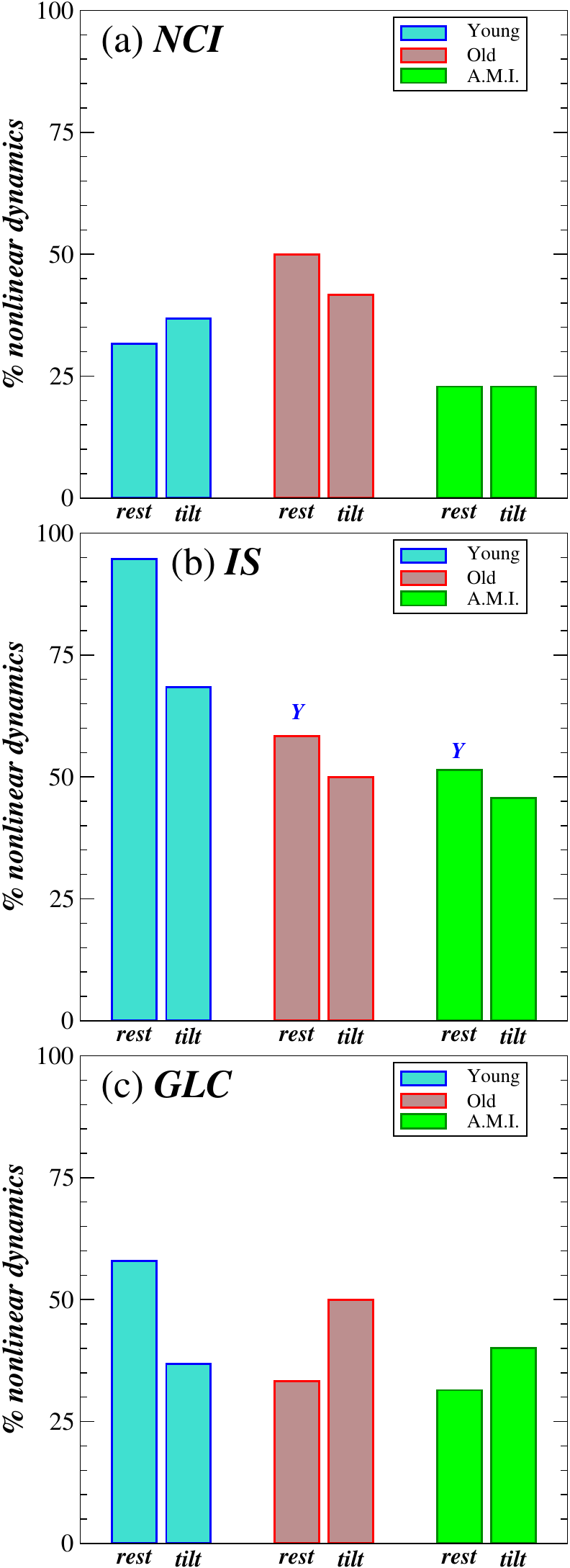}
\caption{\label{fig:NLnum} Percentage of significant nonlinear dynamics obtained counting the subjects for which each nonlinear dynamic measure was significantly different for the original heart period than for the surrogate time series. Results are shown for each of the three proposed methods (a, Normalized Complexity Index; b, Information Storage; c, Gaussian Linear Contrast) applied to Young subjects (blue), Old subjects (red) and post AMI patients (green) in the $rest$ and $tilt$ conditions. Statistical analysis: $\#, p<0.05 rest $ vs. $tilt$; $Y, p<0.05$ vs. Young.}
\end{figure}




\section{Discussion}

The purpose of this study was to perform a comparative investigation of the aptitude of three recently proposed nonlinear dynamic measures to quantify the presence and the extent of nonlinear dynamics in short-term recordings of HRV obtained under different physio-pathological states. In a time series observed as a realization of a stochastic process, nonlinear dynamics are typically described as nonlinear correlations between time-lagged variables taken from the process \cite{theiler1992testing}. In our analysis, these correlations are detected directly in terms of mutual information between the present and the past sample of the process by the information storage (IS), inversely in terms of conditional entropy of the present sample given the past by the normalized complexity index (NCI), or in terms of deviation of the estimated correlation from the value that would be expected in case of linear correlations by the Gaussian linear contrast index (GLC).
Our results document that differences in the detection and quantification of nonlinearity emerge among the three measures, suggesting that a given nonlinear dynamic measure may be more or less sensitive to the detection of specific types of nonlinear dynamics, and that distinct nonlinear dynamic structures may underlie the generation of HRV depending on the physio-pathological condition under analysis.

The opposite variations exhibited by the indexes of conditional entropy and information storage when moving from rest to tilt or while comparing two groups (Fig. 2a,b) can be explained considering that NCI and IS are related to each other as they reflect respectively the unpredictability and the predictability of the dynamics \cite{xiong2017entropy}. The lower NCI and higher IS measured in response to tilt indicates higher predictability of HRV, likely associated to the activation of the sympathetic nervous system induced by the postural challenge \cite{porta2007progressive,faes2014conditional}. Such an activation seems less important in the old and post-AMI groups compared with the young subjects, as documented by the smaller variation of the indexes (though still statistically significant in AMI) and by the higher NCI/lower IS found during tilt in Old and AMI compared to Young. Confirming previous studies \cite{lombardi1987heart,nollo2002evidence}, these results suggest that aging and myocardial infarction are associated with higher sympathetic tone and reduced capability to cope with the postural challenge with further sympathetic activation.

On the other hand, the trends displayed by the GLC measure (Fig. 2c) are in agreement with those of the conditional entropy in the young subjects (both GLC and NCI decrease with tilt), and with those of the information storage in the AMI patients (both GLC and IS increase with tilt).
The different behavior of the GLC index can be explained by considering that the GLC index reflects the extent to which the correlations of the time series deviate from those expected in the linear Gaussian case \cite{carpena2019transforming}, and thus it is not dependent on the extent of linear correlations within the observed time series. As such, GLC should be interpreted as a direct measure of nonlinearity rather than as a regularity index. This is confirmed by the consistent changes between conditions displayed by the absolute values of GLC and by the difference between the index and the median value of its surrogate counterparts (Fig. 2c vs. Fig. 3c). On the contrary, IS is a regularity measure which accounts for both linear and nonlinear correlations, and its increase with tilt is mainly driven by the enhancement of linear HRV correlations. In fact, when the effects of linear correlations are removed by subtracting the median on the surrogates the behavior of IS becomes more similar to that of GLC (Fig. 3b,c).

The quantification of nonlinear HRV dynamics based on computing the deviation of each measure from the median level of its surrogate distribution reveals that, in the young healthy subjects, the transition from rest to tilt is associated with a decreased degree of nonlinearity (Fig. 3a,b,c). This result is in agreement with the observation that nonlinear dynamics are reduced in the presence of an increased sympathetic activity \cite{porta2000prediction,porta2007complexity}. While this latter finding is observed consistently for all three measures, other behaviors characterizing the AMI group are detected peculiarly by individual indexes. In particular, in the AMI patients the information storage was associated to a reduced importance of nonlinear HRV dynamics at rest (Fig. 3b). Again, this result may be related to the sympathetic overactivity characterizing the post-infarction phase \cite{lombardi1987heart}. Another peculiar result is the increased contribution of nonlinear dynamics to HRV measured during tilt in the post-AMI patients (Figs. 2c, 3c). This finding is novel and unexpected, and may reveal that a distinct type of nonlinearity takes place when the orthostatic stress is delivered in the presence of higher sympathetic tone. 

In spite of the similar trends observed for the absolute values and for the deviation from the surrogate median value of NCI and IS (Fig. 3a,b), the two information measures exhibit different percentage of significant nonlinearity in the various conditions (Fig. 4a,b). In agreement with previous studies assessing complexity through prediction measures \cite{porta2000prediction,porta2007complexity}, the amount of nonlinear dynamics detected by the complexity measure based on local sample entropy was small in the young healthy subject at rest, and did not change significantly with the sympathetic activation induced by tilt or related to age and pathology (Fig. 4a). On the contrary, using a regularity measure based on information storage nonlinear dynamics were found consistently in the young subjects at rest, and their incidence decreased significantly with postural stress and in the old and post-AMI groups (Fig. 4b). This finding may reflect the fact that the spontaneous cardiovascular regulation occurs through a variety of nonlinear mechanisms (e.g., saturation of receptors, effects of the respiratory centers at the brain stem level,  interaction between sympathetic and parasympathetic nervous systems, etc.) in resting conditions \cite{koepchen1991physiology}, and the rise of a specific oscillatory component (i.e., the low frequency one related to sympathetic activation) tends to simplify the dynamics reducing nonlinear components. The reduction of nonlinear dynamics with the tilt-induced sympathetic activation is confirmed (though to a lower extent) by the test using the GLC measure (Fig. 4c). The same test however indicates a tendency to increase the rate of detection of nonlinear dynamics with tilt in the old subjects and AMI patients. This could suggest that mechanisms more complex  than a pure sympathetic activation are triggered by the orthostatic stress delivered in the elderly and pathological states \cite{luukinen2004orthostatic}.

However, more methodological factors might be responsible for the disparity of the conclusions drawn by the exploited markers. In a previous study \cite{porta2015complexity}, different conclusions about HRV nonlinear dynamics were drawn using different nonlinearity measures (based on nonlinear prediction and time irreversibility) in fetal HRV recordings as well as in adults during graded head-up tilt. In particular, the different responses to tilt documented by Porta et. al \cite{porta2015complexity} using nonlinear prediction and time irreversibility are comparable to those observed here using the IS and GLC indexes. 
While the different rates of detection of nonlinearity were explained in \cite{porta2015complexity} in terms of the different time scales spanned by the measures employed, this interpretation should not hold in our case since all measures work in the same low dimensional embedding space (m=2 in this study).
A difference between the information approach and the Gaussian Linear contrast method is that NCI and IS are obtained aggregating all time lags in the computation of the measure, while GLC results from analyses performed individually for each lag and then aggregated in the final measure. In addition, the performance of GLC might be affected by the comparison with surrogates that might have amplified eventual residual departures from gaussianity present in the surrogate data due to finite size effects.
As to the differences observed between NCI and IS, they might be related to the fact that the information storage is a direct measure of nonlinear correlations expressed in terms of mutual information, while the conditional entropy measure reflects not only nonlinear correlations but also the entropy of the observed time series; moreover, the different coarse graining approaches underlying NCI and IS using, respectively, equal versus different cell size \cite{porta2007complexity}, and the dependence of the cell size on the parameters set for the analysis (i.e., respectively, tolerance and number of nearest neighbors) might have played a role.
In order to better elucidate the nature of the observed differences and the capability of the various measures to detect different types of nonlinear dynamics, future studies should consider extension of these measures where longer temporal scales can be explored (e.g., analyzing longer stationary recordings and/or employing methods for dimensionality reduction \cite{faes2014conditional}), and deviations of the estimator specific parameters from their nominal typical value are investigated (e.g., for the information measure, the parameter setting the size of the cell used in the multidimensional space to estimate probabilities \cite{richman2000physiological,faes2015estimating}).

\begin{acknowledgments}
P.C. and P.B.G. acknowledge financial support  by the Consejer\'{\i}a de Conocimiento, Investigaci\'on y Universidad, Junta de Andalucía  and European Regional Development Fund (ERDF), ref. SOMM17/6105/UGR, FQM-362, and FQM-7964
\end{acknowledgments}



\bibliography{entropy-bib}
\bibliographystyle{plain}
\end{document}